\documentclass{article}

\usepackage{amssymb}
\usepackage{amsmath}
\usepackage[dvips]{graphicx}
\begin{document}


\begin{center}\textbf{\large``Skew'' scattering of cold unpolarized neutrons in
ferromagnetic crystal}

\bigskip

O.G. Udalov \footnote[1]{udalov@ipm.sci-nnov.ru}

\smallskip

\small Institute for physics of microstructures RAS, (603950)
GSP-105, Nizhny Novgorod, Russia

\bigskip

\end{center}

\noindent \textbf{Abstract}

\noindent The problem of neutron scattering by the single magnetic atom is theoretically considered in the second order perturbation theory. It is demonstrated that elastic scattering of unpolarized neutron by magnetic atom is skewed, i.e. contains the term with the symmetry of mixed product of the atom magnetic moment and wave vectors of incident and scattered neutrons $([\vec{k}\times \vec{k}']\cdot \vec{h})$. The problem of dynamical diffraction of unpolarized neutrons by the perfect ferromagnetic crystal is investigated. The case is considered when the Bragg condition is satisfied for two reciprocal lattice vectors. In this case neutron skew scattering manifests itself as the dependence of the diffracted beam intensity on the sign of the crystal magnetization. The calculations have been done for the diffraction of unpolarized neutrons by Dy crystal. In case of the diffraction by Dy the change of the intensity under the magnetization reversal achieves 50\%.

\noindent \textbf{}

\noindent

\noindent \textbf{Introduction}

As it is well known a neutron is the neutral particle and there are no spin-orbit interaction and Lorenz force for it. This leads to absence of Hall-like effects for neutrons moving in uniform magnetic field. Magnetic field influences on neutron behavior due to existence of a magnetic moment of the particle. Magnetic dipole interaction is described by the operator $-\mu (\hat{\vec{\sigma }}\vec{B})$ [1], where $\mu $ is the neutron magnetic moment, and $\vec{B}$ is the induction of magnetic field, $\hat{\vec{\sigma }}$ is Pauli matrices vector. It is well known that the interaction of this kind leads to Hall-like effect in systems with non-uniform magnetic field. This phenomenon, so-called topological Hall Effect (THE), has been theoretically predicted [2,3] and experimentally observed [4] for conduction electrons in solids. In this paper effect similar to the topological Hall Effect is theoretically investigated for neutrons.

In the case of electrons Pauli term describes the exchange interaction of conduction electrons with localized ones. It has the form $-J(\hat{\vec{\sigma }}\vec{M})$, where $J$ is an exchange constant, $\vec{M}$ is the unit vector co directed with magnetization. This term can be reduced to the effect of fictitious Lorenz force which is non zero only in the systems with non-coplanar and at least two-dimensional magnetization spatial distributions. It is naturally to expect appearance of the effect for neutrons similar to THE in the same magnetic systems. However in contrast to the exchange interaction depending on the magnetization of the media the magnetic dipole interaction of neutrons contains the induction of the magnetic field. Therefore exactly the magnetic induction should satisfy the criteria mentioned above. The simplest example of three dimensional non-coplanar magnetic induction spatial distributions is the field of the point dipole. The scattering of unpolarized neutrons by a magnetic dipole in the first order Born approximation does not show any Hall-like effects [1]. But beyond the first order of the perturbation theory new effects appear. Some kind of Hall Effect for polarized neutrons was firstly described by Maleyev and Toperverg [5,]. It has been shown that inelastic scattering cross section of polarized neutrons contains the term $(\vec{P}[\vec{k}\times \vec{k}'])$, where $\vec{P}$ is the polarization of incident neutron beam, $\vec{k}$, $\vec{k}'$ are the wave vectors of the incident and scattered beams. This effect was observed experimentally by Okorokov, Gusakov, Otchik and Runov [6]. In the present paper it is shown that the elastic scattering cross section of unpolarized neutrons by a single magnetic dipole can contain the term $(\vec{S}[\vec{k}\times \vec{k}'])$, where $\vec{S}$ is the dipole magnetic moment. This term also arises in the second order perturbation theory. Such a peculiarity of cross section can be named as ``neutron skew scattering'' by analogy with skew scattering of electrons in ferromagnets [7].

In the ferromagnetic crystals magnetic dipoles form the regular lattice. The peculiar properties of the neutron scattering by the single dipole should reveal itself in the diffraction by such a crystal. In the present paper the dynamical diffraction of unpolarized neutrons by the perfect ferromagnetic crystal is investigated. In contrast to previous works on this subject [8-10 and many other] ``three wave'' approximation is considered, i.e. the approximation in which the Bragg condition is nearly satisfied for two reciprocal lattice vectors. Only in this case non-coplanar character of the microscopic magnetic field plays the role and the neutron ``skew'' scattering makes the contribution $\propto (\vec{M}[\vec{k}\times \vec{k}'])$ to the diffraction ($\vec{M}$ is the crystal magnetization).

The paper is organized as follows. In the second section the neutron cross section by a single magnetic atom is calculated in the frame of perturbation theory. In the third section the unpolarized neutron diffraction by a perfect ferromagnetic crystal is considered.

\textbf{}

\noindent \textbf{Magnetic elastic scattering of unpolarized neutron by point magnetic dipole}

Consider cold neutron magnetic scattering by an atom. Spin
dependent scattering amplitude of such a process is given by the
formula [11]
\begin{equation} \label{GrindEQ__1_}
\hat{f}^{(1)} (\vec{k},\vec{k}')=2r_{0} \gamma SP(\vec{k}-\vec{k}')(\hat{\vec{\sigma }}\cdot \vec{q}_{m} ),
\end{equation}
\begin{equation} \label{GrindEQ__2_}
\vec{q}_{m} =\vec{h}-(\vec{h}\cdot \vec{e})\vec{e},
\end{equation}
\begin{equation} \label{GrindEQ__3_}
\vec{e}=(\vec{k}-\vec{k}')/|\vec{k}-\vec{k}'|.
\end{equation}
Here $\hat{f}^{1} $ is a spin matrix, superscript (1) means that
the scattering amplitude is calculated in the first order of
perturbation theory, $r_{0} $ is classical electron radius,
$\gamma =-1.913$ is a neutron magnetic moment expressed in the
units of nuclear magneton, $S$ is the atom spin (spin of the
electrons), $P(\vec{k}-\vec{k}')$ is a magnetic form factor,
$\vec{h}$ is the unit vector along the magnetic moment of atom. In
the sake of simplicity lets consider the point dipole, i.e. assume
$P(\vec{k}-\vec{k}')=1$. The cross section of unpolarized neutron
elastic magnetic scattering $\sigma _{m}^{(1)} (\vec{k},\vec{k}')$
in the first order Born approximation is given by the well known
formula [1]:
\begin{equation} \label{GrindEQ__4_}
\sigma _{m}^{(1)} (\vec{k},\vec{k}')=(r_{0} \gamma )^{2} S^{2} (1-(\vec{e}\vec{h})^{2} )
\end{equation}
The scattering is reciprocal
\begin{equation} \label{GrindEQ__5_}
\sigma _{m}^{(1)} (\vec{k},\vec{k}')=\sigma _{m}^{(1)} (-\vec{k}',-\vec{k})
\end{equation}
and does not contain term $(\vec{h}[\vec{k}\times \vec{k}'])$.

In the second order Born approximation the scattering amplitude has the form
\begin{equation} \label{GrindEQ__6_}
\hat{f}^{(2)} (\vec{k},\vec{k}')=\int \frac{\hat{V}_{m} (\vec{k}'-\vec{q})\hat{V}_{m} (\vec{q}-\vec{k})}{\varepsilon -\hbar ^{2} q^{2} /2m_{n} +i\delta } d^{3} q
\end{equation}
\begin{equation} \label{GrindEQ__7_}
\hat{V}_{m} (\vec{k}-\vec{k}')=-\frac{4\pi \hbar ^{2} }{m_{n} } S(\hat{\vec{\sigma }}\cdot \vec{q}_{m} )
\end{equation}
$\varepsilon =\hbar ^{2} \vec{k}^{2} /2m_{n} $ is neutron energy, $m_{n} $ is neutron mass. This scattering amplitude corresponds to the double neutron scattering. Correction to the scattering cross section of unpolarized neutron () is determined by the interference of the first and second order of perturbation series for scattering amplitude.
\begin{equation} \label{GrindEQ__8_}
\sigma _{m}^{(2)} (\vec{k},\vec{k}')=2TrRe(\hat{f}^{(1)} (\vec{k},\vec{k}')^{*T} \hat{f}^{(2)} (\vec{k},\vec{k}')).
\end{equation}
$Tr$ means the trace over spin coordinates, $Re$means real part. It is easy to show that
\begin{equation} \label{GrindEQ__9_}
Tr(\hat{V}_{m} (\vec{k}'-\vec{k})^{*T} \hat{V}_{m} (\vec{k}'-\vec{q})\hat{V}_{m} (\vec{q}-\vec{k}))=i(\vec{q}_{m1} [\vec{q}_{m2} \times \vec{q}_{m3} ]),
\end{equation}
\begin{equation} \label{GrindEQ__10_}
\vec{q}_{m1} =\vec{h}-(\vec{h}\cdot \vec{e}_{1} )\vec{e}_{1} ,\mathop{}\nolimits^{} \vec{e}_{1} =(\vec{k}-\vec{k}')/|\vec{k}-\vec{k}'|,
\end{equation}
\begin{equation} \label{GrindEQ__11_}
\vec{q}_{m2} =\vec{h}-(\vec{h}\cdot \vec{e}_{2} )\vec{e}_{2} ,\mathop{}\nolimits^{} \vec{e}_{2} =(\vec{k}'-\vec{q})/|\vec{k}'-\vec{q}|,
\end{equation}
\begin{equation} \label{GrindEQ__12_}
\vec{q}_{m3} =\vec{h}-(\vec{h}\cdot \vec{e}_{3} )\vec{e}_{3} ,\mathop{}\nolimits^{} \vec{e}_{3} =(\vec{q}-\vec{k})/|\vec{q}-\vec{k}|.
\end{equation}
Correspondingly
\begin{equation} \label{GrindEQ__13_}
\begin{array}{l} {\sigma _{m}^{(2)} (\vec{k},\vec{k}')=\frac{4\pi \hbar ^{2} }{m_{n} } (r_{0} \gamma )^{3} S^{3} \int \delta (\varepsilon -\hbar ^{2} q^{2} /2m_{n} )(\vec{q}_{m1} [\vec{q}_{m2} \times \vec{q}_{m3} ])d^{3} q ,} \\  \end{array}
\end{equation}
In the coplanar situation, i.e. when $(\vec{h}\cdot [\vec{k}\times
\vec{k}'])=0$, there are no second order correction to the cross
section $\sigma _{m}^{(2)} (\vec{k},\vec{k}')=0$. Indeed the
expression (8) in this case transforms to
\begin{equation} \label{GrindEQ__14_}
\sigma _{m}^{(2)} (\vec{k},\vec{k}')\propto \left([\vec{h}\times (\vec{k}-\vec{k}')]\cdot \vec{F}\right),
\end{equation}
\begin{equation} \label{GrindEQ__15_}
\vec{F}=\int \left\{(\vec{h}\cdot \vec{e}_{2} )(\vec{h}\cdot \vec{e}_{1} )+(\vec{h}\cdot \vec{e}_{2} )(\vec{h}\cdot \vec{e}_{3} )+(\vec{h}\cdot \vec{e}_{3} )(\vec{h}\cdot \vec{e}_{1} )\right\}\delta (\varepsilon -\hbar ^{2} q^{2} /2m_{n} )d^{3} q .
\end{equation}
Let's $\vec{k}$, $\vec{k}'$ and $\vec{h}$ lays in the plane
$(x,y)$, then only z component of the vector $\vec{F}$ gives
contribution to $\sigma _{m}^{(2)} (\vec{k},\vec{k}')$. The z
component of  $\vec{F}$ is zero since the products $(\vec{h}\cdot
\vec{e}_{i} )$ depend only on $q_{z}^{2} $, and  do not depend on
$q_{z} $.

For the case of general orientation of $\vec{k}$, $\vec{k}'$ and
$\vec{h}$ the contribution $\sigma _{m}^{(2)} (\vec{k},\vec{k}')$
to the scattering cross section is an odd function of $\vec{h}$.
Also $\sigma _{m}^{(2)} (\vec{k},\vec{k}')$ changes the sign under
permutation $\vec{k}\to \vec{k}'$, $\vec{k}'\to \vec{k}$ and under
permutation $\vec{k}\to -\vec{k}'$, $\vec{k}'\to -\vec{k}$. Thus
the term $\sigma _{m}^{(2)} (\vec{k},\vec{k}')$ has the same
properties as the mixed product $(\vec{h}\cdot [\vec{k}\times
\vec{k}'])$.

The dependence of the unpolarized neutron scattering cross section
($\sigma _{m}^{} (\vec{k},\vec{k}')=\sigma _{m}^{(1)}
(\vec{k},\vec{k}')+\sigma _{m}^{(2)} (\vec{k},\vec{k}')$) on the
angle between wave vectors of incident and scattered neutrons is
presented. Vectors $\vec{k}$ and $\vec{k}'$ are perpendicular to
the magnetic moment of point dipole. For better representation the
case is presented of unfeasibly strong magnetic interaction at
which $\sigma _{m}^{(2)} (\vec{k},\vec{k}')$ is of order of
$\sigma _{m}^{(1)} (\vec{k},\vec{k}')$. For real magnetic atoms
and wavelengths of cold neutrons the contribution of the second
order is very small $\sigma _{m}^{(2)} (\vec{k},\vec{k}')/\sigma
_{m}^{(1)} (\vec{k},\vec{k}')\propto r_{0} \gamma /\lambda \approx
10^{-4} $ ($\lambda $ is neutron wavelength).

\noindent \includegraphics*[width=3.03in, height=2.89in,
keepaspectratio=false]{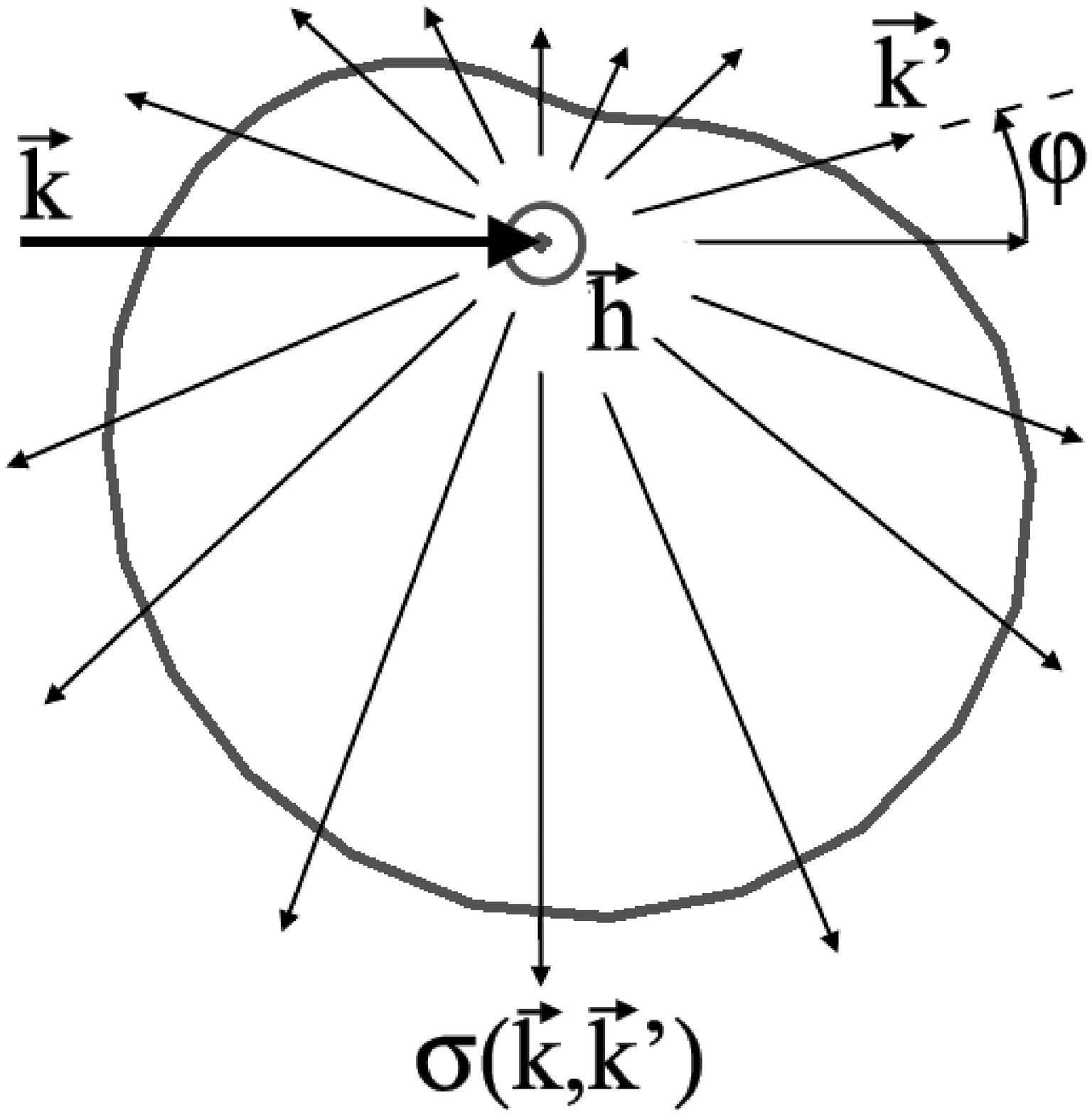}\includegraphics*[width=3.03in,
height=2.89in, keepaspectratio=false]{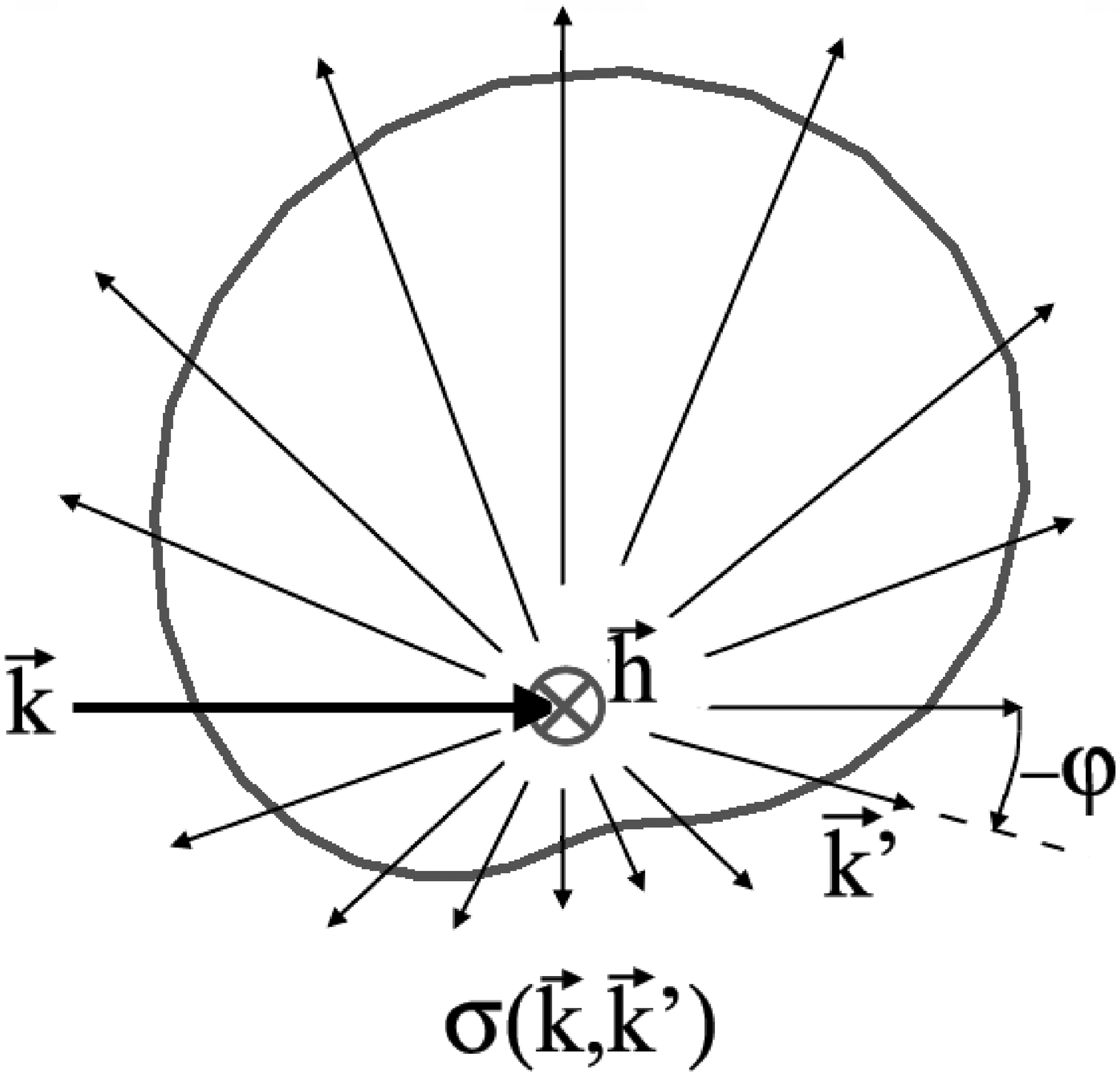}
\begin{center}
   (a)\rule{7cm}{0pt}(b)

\textbf{}

Fig. 1. Dependence of uploarized neutron scattering cross section
(in arbitrary units) on the angle $\varphi $ between wave vectors
of incident ($\vec{k}$) and scattered ($\vec{k}'$) beams. Two
orientations of dipole magnetic moment. Polar plot.
\end{center}

\textbf{}

It can be seen from the Fig. 1 (a) the neutrons scatter mainly to
the right with respect to incident beam and the scattering is
indeed skewed. If one reverse the direction if the magnetic dipole
the preferred direction of the scattering also is reversed with
respect to incident vector wave vector.

\textbf{}

\noindent \textbf{Diffraction of unpolarized neutrons by the perfect ferromagnetic crystal}

 It is reasonable that the phenomena of neutron skew scattering can be enhanced by means of the interference effects. Perfect ferromagnetic crystal represents regular lattice of codirected magnetic dipoles. Let's consider dynamical diffraction of unpolarized neutrons by such a crystal.

 At first analyze the results of the previous section to understand the necessary condition for ``skew'' diffraction of neutrons. As it was shown the ``skew'' scattering appears due to the interference of the once and twice scattered waves (see Fig. 2).

\begin{center} \includegraphics*[width=3.15in, height=2.52in,
keepaspectratio=false]{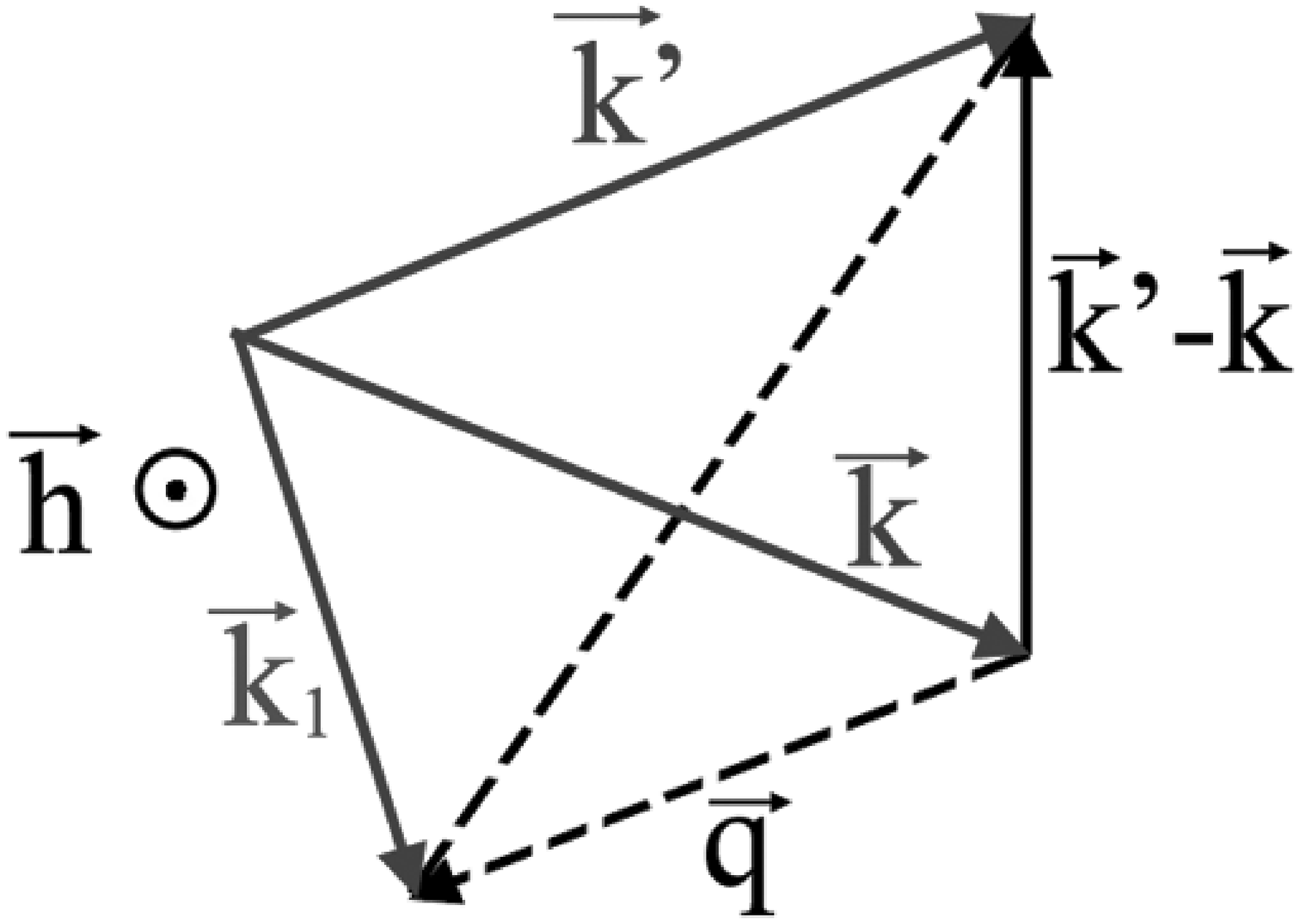}

Fig. 2. Scattering from the state $\vec{k}$ to the state
$\vec{k}'$. Interference of the once and twice scattered waves.

\end{center}

In the case of the scattering by the single dipole all
intermediate states $\vec{q}$ on the energy surface contribute to
the scattering. In the case of diffraction only the certain
direction of scattering are permitted. Therefore possible
intermediate states determined by the Bragg conditions. If the
Bragg condition is fulfilled only for one reciprocal vector (here
this situation named as ``two wave approximation'') it intuitively
seems that the double scattering processes through the
intermediate state (depicted by the dash line on the figure) are
not taken into account. Indeed as it has been shown earlier [8]
the diffraction in this case is reciprocal
\begin{equation} \label{GrindEQ__16_}
I(\vec{k},\vec{k}',\vec{M})=I(\vec{k}',\vec{k},-\vec{M}),
\end{equation}
\begin{equation} \label{GrindEQ__17_}
I(-\vec{k}',-\vec{k},\vec{M})=I(\vec{k},\vec{k}',\vec{M}),
\end{equation}
and therefore does not contain terms of the from $(\vec{M}\cdot
[\vec{k}\times \vec{k}'])$. In the formulas (16), (17) $I$ denotes
the intensity of neutrons diffracted in the direction $\vec{k}'$
($\vec{k}'-\vec{k}=\vec{g}$, $\vec{g}$ is the reciprocal lattice
vector).

 Double scattering through the intermediate state appears if one considers the ``three wave approximation''. In this case the Bragg condition is nearly satisfied for two reciprocal vectors (see Fig. 3).

 From the formula (8)it is seen that the contribution to the ``skew'' scattering is made by the intermediate states with vectors $\vec{q}$ having non-zero projection on the direction of the atom magnetic moment. Thus if all the reciprocal lattice vectors participating in the diffraction process perpendicular to the magnetic moment of the crystal no skew scattering should appear.

  Let's now investigate the problem of ``three wave diffraction'' for unpolarized neutrons. In accordance with Bloch theorem the wave function of neutron inside the ferromagnet is written as follows
\begin{equation} \label{GrindEQ__18_}
\Psi _{\vec{k}} =\sum _{i}\Psi _{gi}^{k} e^{i(\vec{k}+\vec{g}_{i} )}
\end{equation}
Here $\vec{k}$ is quasimomentum of a neutron in media, $\vec{g}_{i} $ are the reciprocal lattice vectors, $\Psi _{gi}^{k} $ are the two component amplitudes. The system of equations for $\Psi _{gi}^{k} $ has the form
\begin{equation} \label{GrindEQ__19_}
(1-k^{2} /\chi ^{2} )\Psi _{gi}^{k} -\sum _{k}(\hat{N}_{gk} +\hat{M}_{gk} )\Psi _{gi-gk}^{k}  =0
\end{equation}
Neutron energy $\varepsilon $ is related with$\chi $ by the formula$\varepsilon =\hbar ^{2} \chi ^{2} /2m_{n} $. $\hat{N}_{gi} $ are the coefficients of the Fourier series of the nuclear potential.
\begin{equation} \label{GrindEQ__20_}
\hat{N}_{gk} =\frac{2m_{n} N_{c} }{\hbar ^{2} \chi ^{2} } \int _{unit^{} cell}d^{3} r\hat{V}_{nuc} (\vec{r})\exp (i\vec{g}_{k} \vec{r}) .
\end{equation}
$N_{c} $ denotes the number of unit cells per unit of the crystal volume, $\hat{V}_{nuc} (\vec{r})$ is the nuclear potential. In the case of slow neutrons Fermi pseudo potential approximation can be used for the description of neutron-nuclear interaction.
\begin{equation} \label{GrindEQ__21_}
\hat{V}_{nuc} (\vec{r})=\frac{2\pi \hbar ^{2} }{m_{n} } \sum _{i}b\delta (\vec{r}-\vec{r}_{i} )
\end{equation}
Summation in the formula is made over all atoms in a crystal. $b$ is the coherent nuclear scattering length. Here the dependence of the potential on the nuclear spin is not taken into account since the temperature is assumed to be not very low.

\noindent Thus
\begin{equation} \label{GrindEQ__22_}
\hat{N}_{gk} =\frac{4\pi N_{c} }{\chi ^{2} } \sum _{l}b_{l} \exp (i\vec{g}_{k} \vec{R}_{l} ) ,
\end{equation}
where summation is carried over the atoms in one cell, $\vec{R}_{l} $ is the position of atoms.

 $\hat{M}_{gk} $ are the coefficients of the Fourier series of magnetic dipole interaction.
\begin{equation} \label{GrindEQ__23_}
\hat{M}_{gk} =\frac{4\pi N_{c} }{\chi ^{2} } (\hat{\vec{\sigma }}\cdot \vec{q}_{gk} )\sum _{l}p_{l} \exp (i\vec{g}_{k} \vec{R}_{l} )
\end{equation}

\begin{equation}\label{GrindEQ__24_}
\vec{q}_{gk} =\vec{h}-(\vec{h}\cdot \vec{e}_{gk} )\vec{e}_{gk}
,\mathop{}\nolimits^{} \vec{e}_{gk} =\vec{g}_{k} /|\vec{g}_{k} |
\end{equation}
\noindent $p_{l} $ denotes magnetic form factor of the $l$-th
atom. In the ``three wave approximation'' vectors $\vec{g}_{1} $
and $\vec{g}_{2} $ meet the Bragg condition and correspondingly
amplitudes$\Psi _{0}^{k} $, $\Psi _{g1}^{k} $ and $\Psi _{g2}^{k}
$ are non-zero. One should solve the following system of equations
to find them
\begin{equation}\label{GrindEQ__25_}
\left\{\begin{array}{l} {(1-k^{2} /\chi ^{2} -\hat{N}_{0} -\alpha
_{0} (\hat{\vec{\sigma }}\cdot \vec{q}_{0} ))\Psi _{0}
-(\hat{N}_{g1} +\alpha _{1} (\hat{\vec{\sigma }}\cdot \vec{q}_{g1}
))\Psi _{g1} -(\hat{N}_{g2} +\alpha _{2} (\hat{\vec{\sigma }}\cdot
\vec{q}_{g2} ))\Psi _{g2} =0,} \\ {-(\hat{N}_{g1} +\alpha _{0}
(\hat{\vec{\sigma }}\cdot \vec{q}_{g1} ))\Psi _{g0}
+(1-(\vec{k}+\vec{g}_{1} )^{2} /\chi ^{2} -\hat{N}_{0} -\alpha
_{1} (\hat{\vec{\sigma }}\cdot \vec{q}_{0} ))\Psi _{g1}
-(\hat{N}_{g3} +\alpha _{3} (\hat{\vec{\sigma }}\cdot \vec{q}_{g3}
))\Psi _{g2} =0} \\ {-(\hat{N}_{g2} +\alpha _{2} (\hat{\vec{\sigma
}}\cdot \vec{q}_{g2} ))\Psi _{g0} -(\hat{N}_{g3} +\alpha _{3}
(\hat{\vec{\sigma }}\cdot \vec{q}_{g3} ))\Psi _{g1}
+(1-(\vec{k}+\vec{g}_{2} )^{2} /\chi ^{2} -\hat{N}_{0} -\alpha
_{0} (\hat{\vec{\sigma }}\cdot \vec{q}_{0} ))\Psi _{g2} =0}
\end{array}\right.
\end{equation}

\noindent Here the spin dependent part of neutron-atom interaction is presented in the detailed form. This equations contain four magnetic vectors $\vec{q}_{0-3} $ ($\vec{q}_{3} =\vec{h}-(\vec{h}\cdot \vec{e}_{g3} )\vec{e}_{g3} $, $\vec{e}_{g3} =(\vec{g}_{1} -\vec{g}_{2} )/|\vec{g}_{1} -\vec{g}_{1} |$). For simplicity propose that $\vec{g}_{1} $ is perpendicular to magnetization $\vec{g}_{1} \bot \vec{h}$, then $\vec{q}_{0} =\vec{q}_{1} =\vec{h}$, and there are three different magnetic vectors.

Let's consider the symmetry of the equations (25) with respect to
magnetization reversal, i.e. to the replacement $\vec{h}\to
-\vec{h}$. Denote this operation by $\hat{T}$. Terms connected
with magnetic interaction changes their sign under the reversal .
\begin{equation} \label{GrindEQ__26_}
\hat{T}\vec{q}_{gi} =-\vec{q}_{gi} .
\end{equation}
In the case when all three vectors lay in one plane the action of the operator $\hat{T}$ can be compensated by the spin coordinate system rotation by $\pi $ around the axis perpendicular to the vectors $\vec{q}_{0-3} $. It is clear that spin coordinates rotation does not change the intensity of the diffracted beams. Therefore the diffraction of unpolarized neutrons by the perfect ferromagnetic crystal in this case is an even function of magnetization and does not contain the contribution of ``skew'' scattering.

If $\vec{q}_{1-3} $ is non-coplanar the action of the operator $\hat{T}$ can not be compensated by any spin coordinates rotation. Thus the intensity of the diffracted wave can depend on the sing of the magnetization. The condition of non-coplanar alignment of the vectors $\vec{q}_{1-3} $ in case of  $\vec{g}_{1} \bot \vec{h}$ can be formulated in the form
\begin{equation} \label{GrindEQ__27_}
(\vec{g}_{2} \vec{h})^{2} (\vec{h}\cdot [\vec{g}_{1} \times \vec{g}_{2} ])\ne 0.
\end{equation}

 On the base of the equations (25) the diffraction of unpolarized neutrons by Dy crystal has been calculated. The following geometry is chosen. Magnetization of the crystal is parallel the interface (Fig. 3). The unpolarized neutron beam falls onto the crystal surface with the glancing angle $\alpha $. The wave vector of the incident neutrons $\vec{k}_{inc} $ is perpendicular to the crystal magnetization. The vector $\vec{g}_{1} $ is perpendicular to the crystal surface and also to the magnetization. So one of the diffracted wave comes out from the front surface with the wave vector $\vec{k}_{diff} $. The intensity of this diffracted wave is calculated for two direction of the magnetization $\vec{M}$ and $-\vec{M}$. The second reciprocal lattice vector $\vec{g}_{2} $ is tilted with respect to magnetization by the certain angle determined by the Bragg condition.

\begin{center}
\includegraphics*[width=3.72in, height=2.76in,
keepaspectratio=false]{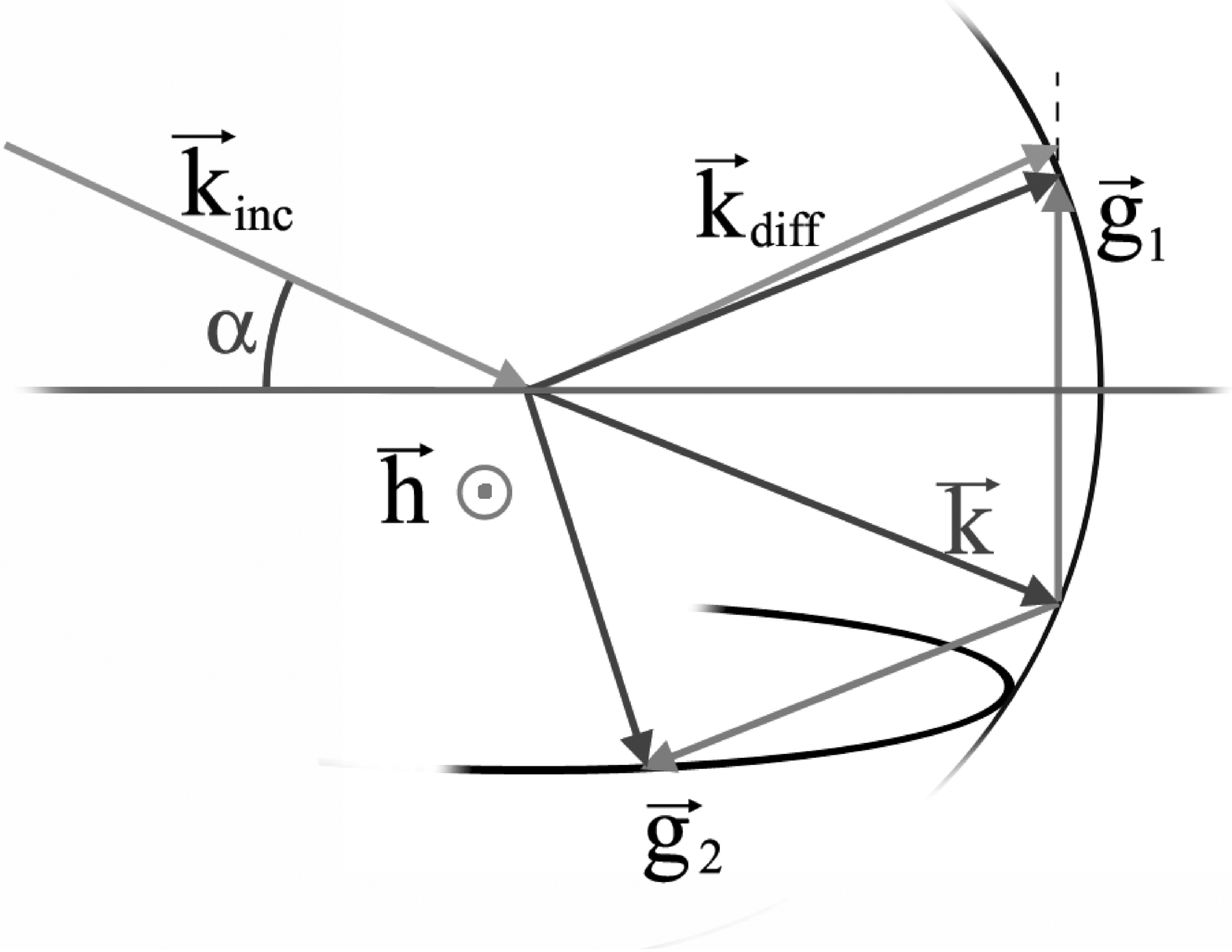}

Fig. 3. ``Three wage'' geometry. Bragg case. $\vec{k}$ is one of
the possible neutron wave vectors inside the crystal.

\end{center}

Analytical solution the system (25) is hardly possible since there
are 12 neutrons waves with the same energy in the crystal, and the
dispersion relation arising from equations (25) is the 12-th order
equation with complicate coefficients. Roots of the dispersion
relation have been found numerically. Using the roots and
corresponding border conditions [8] (the case of semi infinite
crystal has been considered) the intensity of the diffracted waves
has been found. Since the glancing angle is not very small the
intensity of the reflected wave several order smaller than the
diffracted one.

Dy has the simple hexagonal crystal structure. Vector $\vec{g}_{1} $ is codirected with c-axis, and $\vec{g}_{2} $ is along a-axis. The following parameters of the crystal has been used: magnetic moment per atom $\mu =10.8$ $\mu _{nuc} $, the coherent scattering length $b=-1.4\cdot 10^{-12} $ cm, lattice parameters $a=0.35$ nm, $c=$ 0.45 nm. The wave length is $\lambda =0.25$ nm.

The dependence of the intensity of the diffracted beam on the glancing angle for two magnetization direction is presented on Fig. 3.

\begin{center} \includegraphics*[width=4.13in, height=2.34in,
keepaspectratio=false]{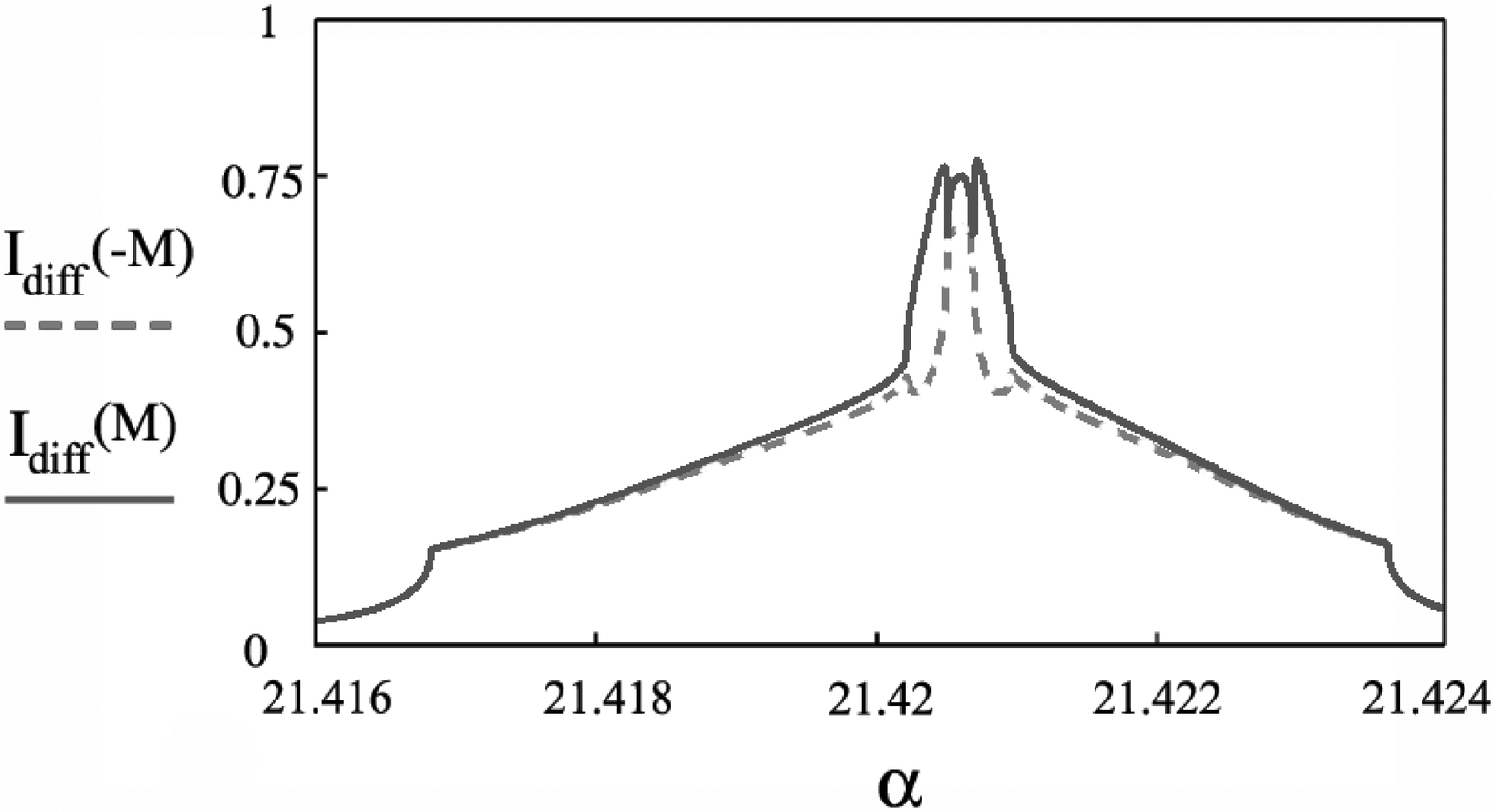}

Fig. 4. The dependence of the intensity of the diffracted beam on
the glancing angle. The intensity is normalized by the intensity
of the incident beam.

\end{center}

As it is seen from the figure the intensity of the diffracted beam depends on the magnetization sign. The relative difference of the intensities for two magnetization directions achieves 50\% in the narrow glancing angle range. Calculations show that the intensity also changes if one interchanges source and detector ($\vec{k}_{inc} \to -\vec{k}_{diff} ,\vec{k}_{diff} \to -\vec{k}_{inc} $) or if one send the incident beam from the back side of the crystal (the replacement $\vec{k}_{inc} \to \vec{k}_{diff} ,\vec{k}_{diff} \to \vec{k}_{inc} $). Thus one can said the intensity depends on the sign of the mixed product $([\vec{k}_{inc} \times \vec{k}_{diff} ]\cdot \vec{h})$. Such dependence is the signature of the ``skew'' scattering of neutrons by a single magnetic atom.

In the considered above geometry (Fig. 3) the incident wave vector is laid in the plane perpendicular to the magnetic field. If the vector comes out of the plane the Bragg condition for the reciprocal lattice vector $\vec{g}_{2} $ can be violated. This leads to disappearing of the effect. Therefore the dependence of the diffracted intensity on the sign of the magnetization arises only in the small solid angle region for the directions of incident beam.

\noindent

\textbf{}

\noindent \textbf{Conclusion}

 Thus in the paper it is demonstrated that elastic scattering of unpolarized neutron by magnetic atom is skewed, i.e. contains the term with the symmetry of $([\vec{k}\times \vec{k}']\cdot \vec{h})$. Such a term in the scattering cross section appears in the second order perturbation theory. In the case of the single atom this contribution to cross section is small and is about $10^{-4} $. Skew scattering of neutrons manifests itself in dynamical diffraction of unpolarized neutrons by perfect ferromagnetic crystal. When the Bragg condition is nearly satisfied for two reciprocal lattice vectors the intensity of the diffracted wave depends on the sign of the magnetization. The calculations have been done for the diffraction of unpolarized neutrons by Dy crystal. The change of the intensity under the magnetization reversal achieves 50\% in the small solid angle region for the directions of incident wave vector.

\noindent \textbf{}

\noindent \textbf{}

\noindent \textbf{}

\noindent \textbf{Acknowledgments}

Author thanks Professor A.A. Fraerman for useful discussions. The
research was supported by the ``Dynasty'' foundation.

\textbf{}

\noindent \textbf{References}

\noindent [1] I. I. Gurevich, L. V. Tarasov, \textit{Physics of Low Energy Neutrons}, (Nauka, Moscow, 1965).

\noindent [2] Ya. Aharonov, A. Stern, Phys. Rev. Lett., \textbf{69}, 25, 3593 (1992).

\noindent [3] D. Loss, P.M. Goldbart, A.V. Balatsky, Phys. Rev. Lett., \textbf{65}, 13, 1655 (1990).

\noindent [4] Y. Taguchi, Y. Oohara, H. Yoshizawa, et. al, Science, \textbf{291}, 2573 (2001).

\noindent [5] A.V. Lazuta, S.V. Maleyev, B.P. Toperverg, Sov.
Phys. JETP 48 (2), 386 (1978).

\noindent [6] A.I. Okorokov, A.G. Gusakov, Ya.M. Otchik et al., Phys. Lett. 65A, 60 (1978).

\noindent [7] F.E. Maranzana, Phys. Rev. 160, 421 (1967)

\noindent [8] C. Stassis, J.A. Oberteuffer, Phys. Rev. B, 10, 12, 5192 (1974)

\noindent [9] M.L. Goldberger, F. Seitz, Phys. Rev., 71, 5, 294 (1947)

\noindent [10] A. Zeilinger, T.J. Beatty, Phys. Rev. B, 27, 12, 7239 (1983)

\noindent [11] B.I. Halperin, P.C. Hohenberg, Phys. Rev., 177, 952 (1969)

\noindent

\noindent

\noindent

\end{document}